# Why Not Consider Closed Universes?

Martin White and Douglas Scott

*Center for Particle Astrophysics, University of California*

*Berkeley, CA 94720-7304*

## Abstract

We consider structure formation and cosmic microwave background (CMB) anisotropies in a closed universe, both with and without a cosmological constant. The CMB angular power spectrum and the matter transfer function are presented, along with a discussion of their relative normalization. This represents the first full numerical evolution of density perturbations and anisotropies in a spherical geometry. We extend the likelihood function vs. $\Omega$ from the *COBE* 2-year data to $\Omega \geq 1$. For large $\Omega$ the presence of a very steep rise in the spectrum towards low $\ell$ allows us to put an upper limit of $\Omega \leq 1.5$ (95%CL) for primordial spectra with $n \leq 1$. This compares favorably with existing limits on $\Omega$. We show that there are a range of closed models which are consistent with observational constraints while being even older than the currently popular flat models with a cosmological constant. Future constraints from degree scale CMB data may soon probe this region of parameter space. A derivation of the perturbed Einstein, fluid and Boltzmann equations for open and closed geometries is presented in an appendix.

*Subject headings:* cosmology:theory — cosmic microwave background — large-scale structure of universe

# 1 Introduction

Much attention has recently been paid to structure formation and cosmic microwave background (CMB) anisotropies in open universes (e.g. Sugiyama & Silk 1994, Kamionkowski & Spergel 1994, Kamionkowski, Spergel & Sugiyama 1994, Kamionkowski et al. 1994, Hu, Bunn & Sugiyama 1995, White & Bunn 1995, Górski et al. 1995, Cayón et al. 1995). The idea that $\Omega < 1$ has been seen as perhaps more promising because of the possibility of fluctuation generation in open inflationary models (e.g. Lyth & Stewart 1990, Ratra & Peebles 1994, Bucher, Goldhaber & Turok 1994, Lyth & Woszczyna 1995, Yamamoto, Sasaki & Tanaka 1995, Liddle et al. 1995). In this paper we would like to focus instead on the alternative case of a universe with a spherical spatial geometry, i.e. a closed universe. We will apply some of the results and techniques developed recently in the study of open and flat spatial geometries to the closed case.

Historically, there has been much interest in closed universes (Wheeler 1968, Hawking 1984), since the compact spatial surfaces can make quantum field theory more clearly consistent than in hyperbolic spaces and renders tractable the integrals of quantum gravity. The idea of the need for closure goes back to Einstein (1934), who regarded it as necessary to solve the "problem of inertia", and was postulated as a boundary condition by early workers in the field (see discussion in Misner, Thorne & Wheeler 1970, p. 543, 704). Recently Linde (1995) has shown that it is possible to produce inflationary models which result in closed universes, in which the universe is "created from nothing" (Tryon 1973, Zel'dovich 1981, Zel'dovich & Grishchuk 1984). This is possible because a closed universe has zero total energy, just as it has zero total momentum and total charge (Landau & Lifshitz 1975). An extensive list of references to the early literature on closed universes can be found in Björnsson & Gudmundsson (1995).

The cosmological constant is often reconsidered in times of apparent astrophysical crisis (see Carroll, Press & Turner 1992), and today's Hubble constant versus age discrepancy is no exception. Recent indications that $H_0 \sim 80\,{\rm km\,s^{-1}Mpc^{-1}}$ partially motivated us to consider models with cosmological constant $\Lambda > 0$. As we discuss in §4, it is possible to gain a few Gyrs in such models, although exactly how old the universe can be depends largely on gravitational lens constraints (and also on constraints from large-angle and degree-scale CMB anisotropies). But certainly $\Lambda$ is not the universal panacea for the age problem.

While we allow for a non-zero $\Lambda$, we restrict our attention to those models which start at arbitrarily small scale (the Big Bang). Models without a Big Bang (i.e. having excessive amounts of $\Lambda$) can be excluded on quite general grounds (Ehlers & Rindler 1989). In order to specify an initial spectrum of fluctuations, we shall implicitly assume an early period of inflation which gave rise to scale invariant potential fluctuations (though we do not calculate the spectrum arising from any *particular* inflationary model). Note that we only concern ourselves with scalar perturbations here, neglecting the possibility of tensor modes. We further restrict our attention to the well-motivated case of adiabatic density fluctuations, although many of our general results would apply to isocurvature fluctuations as well, and it would certainly be possible to perform calculations for any specific isocurvature spectrum. We also consider what would happen if the last scattering surface was at the geometrical antipode (a putative "solution" of the smoothness problem), although in reality this situation is inconsistent with current data.

The outline of the paper is as follows. The next section reviews the closed universe geometry and establishes our notation. §§3 and 4 present some material which will be necessary to understand our principle results. The rest of the paper presents the results of our numerical calculations. §5 discusses the CMB fluctuations in a closed universe. Here we perform the first detailed calculations of the anisotropy angular power spectrum in closed models, by accurate integration of the coupled Einstein, fluid and Boltzmann equations (for more detials see Hu et al. 1995, hereafter HSSW, and references therein). An extensive appendix gives a derivation of these equations for arbitrary geometry. §6 reviews the relation to large-scale structure, with explicit calculation of the matter transfer function and accurate calculation of the normalization relative to the CMB, including the effects of growth rate and non-trivial gravitational potential evolution. We show that the continuation to closed geometries is benign. In §7 we use the COBE 2-year data to place an upper limit on $\Omega_{\rm tot}$, and §8 reviews an extreme model suggested by Harrison (1993). §9 reviews other limits on closed universes, and §10 contains our conclusions.

# 2 Spherical Geometry

We will use the following notation for contributions to the critical density $\rho_{\rm crit} \equiv 3H_0^2/(8\pi G)$: $\Omega_m$, $\Omega_r$, $\Omega_\Lambda$, $\Omega_K$, arising from the (pressure free) matter, the radiation (photons plus massless neutrinos), the cosmological constant and the curvature, respectively. The Friedmann equation can be used to relate these $\Omega$'s:

$$1 = \Omega_{m0} + \Omega_{r0} + \Omega_{\Lambda 0} + \Omega_{K0}. \qquad (1)$$

We will generally use these quantities to parameterize the present state of the universe, and hence the subscript '0' will be implicit throughout the paper. We will also sometimes refer to the quantity $\Omega_{\rm tot} = \Omega_m + \Omega_r + \Omega_\Lambda$. The curvature contribution can then be written as $\Omega_K = 1 - \Omega_{\rm tot}$. Note that for a closed universe $\Omega_K < 1$.

The closed universe has metric

$$\begin{aligned} ds^2 &= -dt^2 + a^2(t)\left[\frac{dr^2}{1-Kr^2} + \right. \\ &\quad \left. r^2\left(d\theta^2 + \sin^2\theta\, d\phi^2\right)\right], \end{aligned} \qquad (2)$$

where $K = H_0^2(\Omega_{\rm tot} - 1)$ is the "curvature", and we use units where $c = 1$. $H_0$ is the Hubble constant today, which we will usually denote by $h = H_0/100\,{\rm km\,s}^{-1}{\rm Mpc}^{-1}$. A closed universe has curvature $K > 1$, i.e. a real radius of curvature. The open universe can be obtained by analytic continuation onto the imaginary axis.

The metric can also be written in terms of the development angle $\chi$ and the conformal time $\eta \equiv \int dt/a(t)$, as

$$\begin{aligned} ds^2 &= a^2(\eta)\left[-d\eta^2 + K^{-1}\left\{d\chi^2 + \right.\right. \\ &\quad \left.\left. \sin^2\chi\left(d\theta^2 + \sin^2\theta\, d\phi^2\right)\right\}\right]. \end{aligned} \qquad (3)$$

Intervals along geodesics are then related by $d\chi = \sqrt{K}\,d\eta$. Note that in terms of these coordinates (3) is just the metric of a 3-sphere, with us at the "North" pole. It is sometimes useful to imagine this geometry as a sphere embedded in a 4-dimensional Euclidean space $(w, x, y, z)$ by (Weinberg 1972)

$$\begin{aligned} w &= \cos\chi \\ x &= \sin\chi\cos\theta \\ y &= \sin\chi\sin\theta\cos\phi \\ z &= \sin\chi\sin\theta\sin\phi\, . \end{aligned} \qquad (4)$$

In terms of this picture all null geodesics (light rays) form great circles on this sphere, with a *maximum separation* when $\chi = \pi/2$. A particularly important point in the spherical space is the antipode, $\chi = \pi$, where the extrapolation of any two geodesics arriving at our position will cross. Thus for light rays with fixed angular separation, the coordinate distance subtended increases from zero (at $\chi = 0$) to a maximum (at $\chi = \pi/2$) and then returns to zero again (at $\chi = \pi$). Equivalently, an object placed at the antipode will fill the whole sky. We will see that this structure has important consequences for the angular power spectrum of CMB anisotropies.

## 3 Inflation, Evolution & Flatness

In a closed universe, with $\Lambda = 0$ for the moment,

$$\Omega(\tilde\eta) = \frac{2}{1+\cos\tilde\eta}, \qquad (5)$$

where $\tilde\eta \equiv \sqrt{K}\eta$ is the development angle, which runs from 0 to $2\pi$. This means that if you live in such a universe, you can tell what stage of evolution you are at by measuring $\Omega_m$ alone. The universe evolves from $\Omega = 1$ at the Big Bang, to $\Omega \to \infty$ at maximum expansion, when $\rho_{\rm crit} \propto H^2 \to 0$, and then back to $\Omega = 1$ at the Big Crunch.

One consequence of this is that the universe spends much of its time near $\Omega = 1$. For these $\Lambda = 0$ models then, an argument can be made (A. Linde, private communication) which is a twist on the old Dicke flatness argument (Dicke 1970, Dicke & Peebles 1979). In a closed model with $\Lambda = 0$, the universe has $1 < \Omega < 2$ for half of its total conformal time (this has to be reduced by another factor of 2 if you are only interested in the expanding phase). [In terms of cosmic time $t$ rather than conformal time $\eta$, this is about 18%, or half of that if the contacting phase is neglected.] Hence it is relatively likely that we would find $\Omega$ not far from 1 in a closed model. It could then be argued that if we observe $\Omega \sim 1$, there is no flatness problem *providing* that $\Omega > 1$, and hence we probably live in a closed universe. Of course these are just a posteriori arguments based on one sample, and therefore may be entirely meaningless (see e.g. Evrard & Coles 1995).

On the other hand, as is well known, the flatness problem is exacerbated when $\Lambda \neq 0$ (e.g. Martel & Wasserman 1990), since the value of $\Omega_m$ deviates very rapidly from unity near the present time, when $\Lambda$ dominates the expansion. Thus there is a naturalness problem (why is $\Omega_m \sim \Omega_\Lambda \sim 1$?) even in flat (i.e. $\Omega_m + \Omega_\Lambda = 1$) models with $\Lambda \neq 0$. We shall consider models with $\Omega_\Lambda > 0$ despite these philosophical problems (Weinberg 1989), taking the view that even unlikely regions of possible parameter space deserve some exploration.

We shall implicitly assume throughout this paper that the initial spectrum of density perturbations is scale invariant (in the gravitational potential, see later) and sufficiently small to not overproduce fluctuations in the CMB. Such assumptions can be motivated by inflationary models which result in $\Omega_{\rm tot} > 1$. Self-consistently we will neglect any contribution to the CMB anisotropy from gravitational waves, which should be sub-dominant for scale invariant spectra. Since "standard" inflation generically gives $\Omega_{\rm tot} = 1$, but allowing for a $\Lambda$ component (Peebles 1984, Turner, Steigman & Krauss 1984), we will assume that we can have $\Lambda$ in some $\Omega_{\rm tot} > 1$ models also.

Recently Linde (1995) has shown that it is possible to produce inflationary models which result in closed universes, in which the universe is "created from nothing" (Zel'dovich & Grishchuk 1984). The universe which emerges from nothing, created by a quantum tunneling process, is generically quite homogenous if the tunneling probability is suppressed enough (which selects the most symmetric path as the most probable one). The tunneling thus solves the homogeneity, isotropy and horizon problems before "inflation" begins, allowing for a truncated period of inflation so as not to inflate the curvature radius well outside the horizon. Specific models which lead to closed universes can be constructed (Linde 1995) using scalar field (inflaton) potentials of the form $\phi^2 \exp(\phi^2)$ or by introducing non-minimal couplings of the inflaton field to the curvature $R$. We shall assume for definiteness that inhomogeneities in the bubble which forms our present universe by tunneling are smaller than those produced during the subsequent period of inflation (Linde & Mezhlumian 1995), though this need not necessarily be the case.

## 4  The Classical Tests

A thorough discussion of the classical cosmological tests in a closed universe can be found in e.g. Sandage (1961,1988), Gott, Park & Lee (1989), Durrer & Kovner (1990), Björnsson & Gudmundsson (1995). In this section we remind the reader of some properties of closed universes which will be necessary to understand our later results. Helpful diagrams of $a(t)$ are presented by Felten & Isaacman (1986). Readers who are unfashionable enough to be familiar with closed models can skip this section entirely.

Except in §8, where we consider an extreme model, we will restrict ourselves to the parameter range $(\Omega_m, \Omega_\Lambda) \in [0,2] \times [0,2]$. We will use the notation $\Omega_\Lambda \equiv \Lambda/3H_0^2$ for the effective contribution of the cosmological constant to $\Omega$. The general Friedmann equation is

$$\frac{\dot{a}}{a} = H_0 \sqrt{\Omega_r a^{-4} + \Omega_m a^{-3} + \Omega_K a^{-2} + \Omega_\Lambda}$$
$$\equiv H_0 E(z) , \qquad (6)$$

with $1 + z \equiv a^{-1}$ and $a_0 = 1$. The acceleration equa-

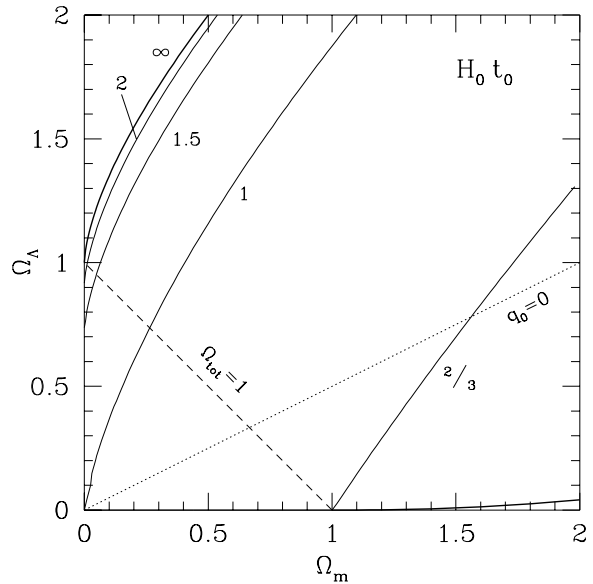

Figure 1: Contours of the age of the universe, in units of the Hubble time $H_0^{-1}$. Also shown (solid line top left) is the critical line for an infinite age universe and (solid line bottom right), a universe which becomes asymptotically static. Universes above (below) these lines have no big bang (have a big crunch). Lines of constant $q_0 = \frac{1}{2}\Omega_m + \Omega_r - \Omega_\Lambda$ are parallel to the $q_0 = 0$ line shown dotted.

tion can be written

$$\frac{\ddot{a}}{a} = -H_0^2 \left( \frac{\Omega_m}{2} a^{-3} + \Omega_r a^{-4} - \Omega_\Lambda \right). \qquad (7)$$

Since $\Omega_r$ is so small today, we make little error by neglecting it for this discussion, although we always include it in numerical calculations.

Using (6), the age integral is

$$H_0 t(z) = \int_z^\infty \frac{dz'}{(1+z')E(z')}, \qquad (8)$$

while the conformal time integral is

$$H_0 \eta(z) = \int_z^\infty \frac{dz'}{E(z')}. \qquad (9)$$

We show contours of $H_0 t_0$ in Fig. 1. As can be seen, increasing $\Omega_\Lambda$ for fixed $\Omega_m$ results in longer ages. This is one of the reasons why cosmological constant models are re-considered periodically (e.g. Gunn & Tinsley 1975, Tinsley 1977, Zel'dovich & Sunyaev, Sandage & Tammann 1984). While the age (in Hubble units) increases rapidly towards the top left of Fig. 1, we will see that much of that region of parameter space is ruled out by classical cosmological tests and by lensing constraints. For a concrete feeling of what these ages mean, we can write

$$t_0 = 12.22 \, \mathrm{Gyr} \, (H_0 t_0) \left(\frac{H_0}{80}\right)^{-1}. \quad (10)$$

By considering the acceleration equation ($\ddot{a} = 0$), we see that there is an inflexion point in the Hubble constant evolution when

$$(1 + z_{\mathrm{loit}})^3 = \frac{2\Omega_m}{\Omega_\Lambda}. \quad (11)$$

This 'loitering' phase can be tuned to be arbitrarily long, giving rise for example to the large ages mentioned above. For models with even modest amounts of loitering one runs into problems with classical cosmological tests which measure volumes. In the limit of extreme loitering, one even "recovers" Olber's paradox.

The age of the universe becomes infinite in a model where $\dot{a} \to 0$ at the inflectional point, or equivalently where $E(z)$ is zero at its extremum. These two conditions together imply

$$27 \, \Omega_m^2 \Omega_\Lambda = 4(\Omega_m + \Omega_\Lambda - 1)^3. \quad (12)$$

This has solution

$$\begin{aligned} \Omega_\Lambda &= 1 - \Omega_m + \frac{3}{2}\Omega_m^{2/3} \times \\ &\quad \left\{ \left(1 - \Omega_m - \sqrt{1 - 2\Omega_m}\right)^{1/3} \right. \\ &\quad \left. + \left(1 - \Omega_m + \sqrt{1 - 2\Omega_m}\right)^{1/3} \right\}, \end{aligned} \quad (13)$$

which can also be written in other forms (see Carroll, Press & Turner 1992). We show this "critical line" in the top left of Fig. 1.

These infinite age universes are also known as Eddington-Lemaître models. The degenerate case $\Omega_m = 0$, $\Omega_\Lambda = 1$ is the de Sitter or Steady State model. Models near to this case have a period in the past where $\dot{a} \simeq 0$, and are therefore known as the "loitering", "coasting" or "hesitating" models. In the past they have also been called Lemaître models, and have long been popular as a means of improving the ability for galaxies to have formed by the present (Lemaître 1931, Rawson-Harris 1969, Brecher & Silk 1969, Tomita 1969, Heath 1977, Occhionero et al. 1980); we shall discuss the issue of growth in loitering models further in §6. A large amount of effort was devoted to these Lemaître models in the early days of high redshift quasars and cosmological number counts (e.g. Shklovskiĭ 1966, Kardashev 1966, McVittie & Stabell 1967, Petrosian, Salpeter & Szekeres 1967, Solheim 1968, Petrosian 1969, Petrosian 1974). Today the *lack* of a pile-up at any redshift rules out our being too close to the Eddington-Lemaître case. Note that Loitering universes have divergent $\eta$ as well as $t$. In fact any "path length" integral diverges, since $E(z) \to 0$ for an arbitrarily long range in $z$ (e.g. the lensing probability shown in Fig. 2).

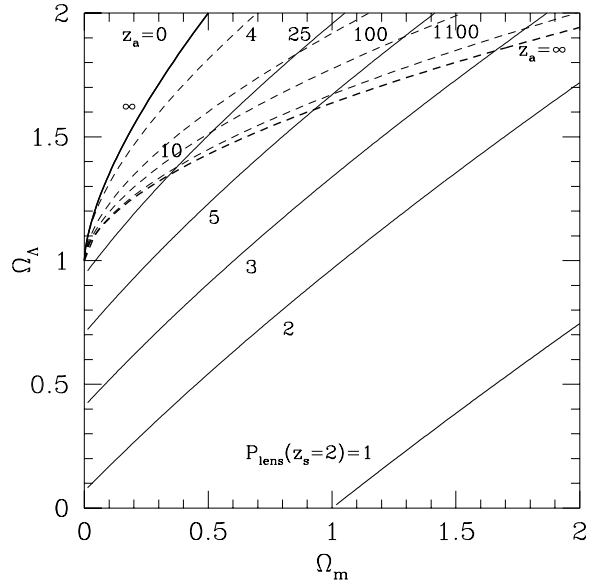

Figure 2: Solid lines show contours of lensing probability, relative to the Einstein-de Sitter model for a source at redshift 2. Dashed lines show the antipode redshift. Both $P_{\mathrm{lens}}(z_s = 2) > 5$ (probably a conservative limit, although estimates vary over $\sim 2$–10) and $z_a < 4$ are ruled out. The lensing constraint is clearly the stronger.

Models with higher $\Omega_\Lambda$ than the critical line have no Big Bang, i.e. they contract from infinity, bounce, and then go into a period of exponential expansion. Such models therefore have a minimum value of $a$, which means a maximum observable redshift. In fact such models must have

$$\Omega_m \leq \frac{2}{z_{\mathrm{max}}^2(z_{\mathrm{max}} + 3)} \quad (14)$$

(Crilly 1968), and since we observe objects out to $z \simeq 5$, while $\Omega_m \gg 0.01$, such models are not allowed (Ehlers & Rindler 1989). Hence the upper left corner of Fig. 1 is definitively ruled out.

Models which have $\Omega_\Lambda = 0$ and $\Omega_m > 1$ will ultimately collapse. However, because of the scale-factor dependence in the Friedmann equation, most models with $\Omega_\Lambda > 0$ will become dominated by the cosmological constant and end up in a de Sitter phase, i.e. they will have an eternally inflating future. There is a small region of parameter space for which $\Omega_m$ wins, and the universe collapses before $\Lambda$ can take over. The critical line is when $\dot{a} \to 0$ for the (future) inflection point $\ddot{a} = 0$. Hence this line is also a solution to (12). Specifically the critical line for collapse is

$$\Omega_\Lambda = 4\Omega_m \left\{ \cos\left[\frac{1}{3}\cos^{-1}\left(\frac{1-\Omega_m}{\Omega_m}\right) + \frac{4\pi}{3}\right]\right\}^3, \qquad (15)$$

for $\Omega_m > 1$ (Felten & Isaacman 1986). This is shown by the solid line in the lower right of Fig. 1.

## 5 The CMB

The postitive curvature of a closed model can shift features in the CMB fluctuation power spectrum to larger angular scales. It has been conjectured for some time (Blanchard 1984, Durrer & Kovner 1990) that this might lead to a reduction in the amplitude of the anisotropies, however no explicit calculations of the anisotropy spectrum have been done.

Early work on CMB anisotropies in non-flat geometries (e.g. Wilson 1983, Tomita & Tanabe 1983, Abbott & Schaefer 1986) concentrated on the large-angle part of the spectrum. In this section we present the first full calculations of the angular power spectrum of fluctuations in a closed universe. The formalism employed is reviewed in the appendix, where the Einstein, fluid and Boltzmann equations for an arbitrary Friedmann-Robertson-Walker geometry are derived. The results were obtained by numerical solution of these coupled equations (see HSSW, and references therein, for more details).

Conventionally one expands the temperatures on the sky in terms of a complete set of functions:

$$T(\theta, \phi) \equiv T_{\rm CMB} \sum_{\ell m} a_{\ell m} Y_{\ell m}(\theta, \phi), \qquad (16)$$

where the spherical harmonics $Y_{\ell m}$ are the curved sky equivalent of Fourier modes, and $T_{\rm CMB}$ is the average temperature of the CMB, so that the $a_{\ell m}$'s are dimensionless. The power spectrum is then defined in terms of the ensemble average of the coefficients:

$$C_\ell \equiv \left\langle |a_{\ell m}|^2 \right\rangle. \qquad (17)$$

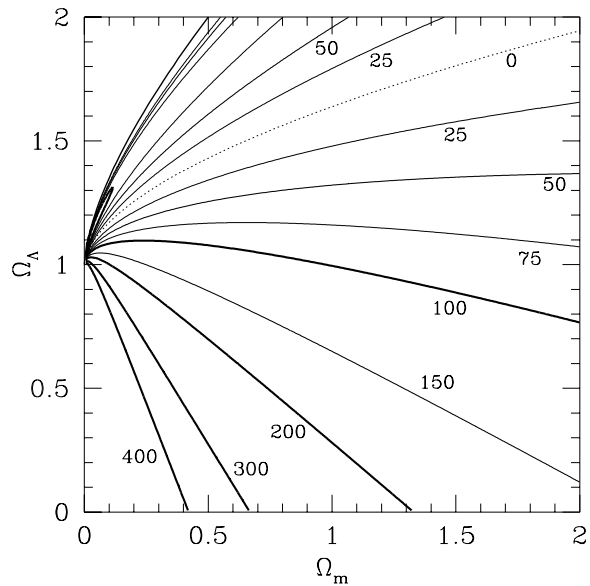

Figure 3: Contours showing the location, $\ell_{\rm peak}$, of the first peak in the CMB anisotopy power spectrum. The other peaks in the power spectrum form an almost harmonic series at $2\ell_{\rm peak}$, $3\ell_{\rm peak}$... Note the region in the upper right of the plot is allowed by classical tests, but will be ruled out by the CMB if the peak is at $\ell_{\rm peak} \sim 200$ as suggested by recent data.

The power per logarithmic interval in $\ell$ is approximately $\ell(\ell+1)C_\ell$, and this vs $\log \ell$ is what we shall refer to henceforth as the angular power spectrum. For reference, multipole $\ell$ probes an angular scale of $\sim 60/\ell$ degrees (see appendix of White et al. 1994) for all but the smallest $\ell$'s.

The most obvious feature of the angular power spectrum is the peak that occurs around $\ell \sim 200$ (see Fig. 4). This peak represents an acoustic oscillation that was maximally overdense at last scattering (for a discussion see Scott & White 1995, Hu, Sugiyama & Silk 1995). Recall that for $\Omega_{\rm tot} \simeq 1$ and $z \gg 1$ the angle subtended by the horizon is roughly $(\Omega_{\rm tot}/z)^{1/2}$ (e.g. Weinberg 1972, eq. [15.5.39]). Since features in the angular spectrum of CMB fluctuations are related to the horizon-size at $z \simeq 1100$, we expect that the whole spectrum will shift $\propto \Omega_{\rm tot}^{-1/2}$. So for example, the main acoustic peak of adiabatic models will be at $\ell_{\rm peak} \simeq 220 \, \Omega_{\rm tot}^{-1/2}$.

One can estimate this effect more precisely by calculating the angular scale subtended by the "sound horizon" at last scattering. The first peak appears at

(Hu & Sugiyama 1995)

$$\ell_{\rm peak} = \pi \left| \frac{r_*}{r_s} \right|, \qquad (18)$$

where, using $\chi_0 - \chi_* = \sqrt{K}\,(\eta_0 - \eta_*)$, we define

$$r_* \equiv \frac{1}{\sqrt{K}} \sin\left[\sqrt{K}\,(\eta_0 - \eta_{LSS})\right], \qquad (19)$$

the effective distance to the last-scattering surface. The sound horizon is

$$r_s(\eta) = \int_0^\eta d\eta\, c_s, \qquad (20)$$

where $c_s$ is the speed of sound. An analytic expression for $r_s$ can be found in Hu & Sugiyama (1995). The higher peaks occur in an almost harmonic series, i.e. $2\ell_{\rm peak}, 3\ell_{\rm peak}, \ldots$

For the case of an open universe, the sine function of (19) is replaced by sinh, which is monotonic in its argument. For closed models, however, there is the possibility of the argument getting bigger than $\pi/2$, or of being $n\pi$ (i.e. an antipode) in which case $r_*$ can have more interesting behavior.

We show in Fig. 3 the position of the first peak as a function of $\Omega_m$ and $\Omega_\Lambda$. Here we have assumed that recombination occured at $z_{\rm rec} = 1100$ for all models, since there is little variation with cosmological parameters. In the numerical work we have followed the recombination process accurately (HSSW).

Notice that towards the top left corner of Fig. 3 the peak moves from high to low $\ell$ and back repeatedly. This is a consequence of the last scattering redshift selecting the equator, the antipode, the equator and so on. In the familiar $\Lambda = 0$ closed models (5), a light ray leaving the Big Bang arrives back at the same point at exactly the Big Crunch ($\chi = 2\pi$). However, this is no longer true for $\Lambda > 0$ closed models, where it is possible to have $\chi \gg 2\pi$, i.e. many antipodes rather than just one. [This is easy to see for an extreme loitering model, where the universe remains static for a long period, allowing light rays to circumnavigate the universe many times.] The peculiar cosmological consequences of an observable antipode have been investigated by many authors (e.g. Petrosian & Salpeter 1968, Rowan-Robinson 1968, Baylis 1970, Biraud & Mavrides 1980, Gott 1985, Gott & Rees 1987, Gott, Park & Lee 1989, Durrer & Kovner 1990, Björnsson & Gudmundsson 1995), although the possibility now seems quite remote. The redshift of the first antipode, $z_a$, was shown in Fig. 2.

The angular power spectra for a variety of models are shown in Figs. 4–6. One can see that the position of the peak moves according to the prescription outlined above. For these models we have specifically assumed that $h = 1$ and $\Omega_B = 0.0125$. The relative heights of the peaks will depend on this choice (higher $\Omega_B$ will enhance the first peak for example) but the positions of the peaks and the large angle (small $\ell$) anisotropy spectrum will not.

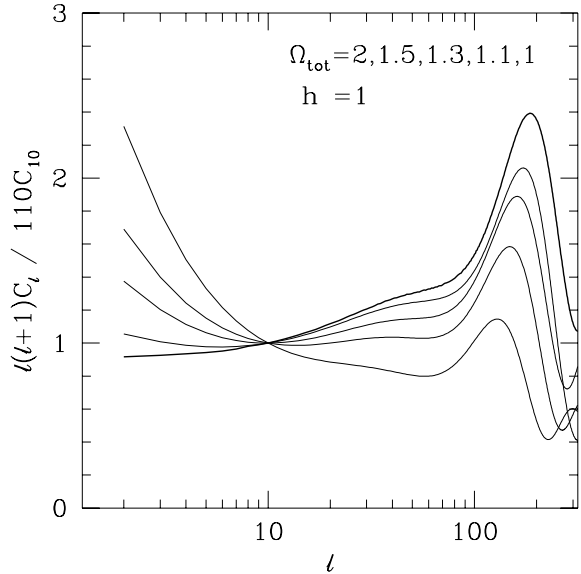

Figure 4: Angular power spectra of CMB anisotropies for models with $\Omega_m > 1$. The position of the first peak moves right with decreasing $\Omega_m$. We have chosen $h = 1$ and $\Omega_B h^2 = 0.0125$ for simplicity. Note the steep rise of the spectrum towards low $\ell$, which allows us to rule out high $\Omega$ models.

In Figs. 4–6, we have assumed that the primordial fluctuation spectrum is "scale invariant", or that the potential fluctuations, $\Phi$, are constant per logarithmic interval in wavenumber (see appendix). Such a spectrum is more physically motivated (e.g. from inflation) than assuming the fluctuations are a power-law in wavenumber, since in a non-flat geometry there *is* an important scale in the problem: the curvature scale. Moreover, such a primordial spectrum leads to smaller deviations from flat CMB spectra than the corresponding power-law. We show the primordial spectra vs. wavenumber in Fig. 7, where one can see that the large scale power is reduced (over the power-law case) in a closed model and enhanced in an open model, counteracting the geometric tendencies to have more (less) large angle power in closed (open) models. The apparent divergence of $P_{\rm prim}$ at small wavenumbers in the open model is counteracted

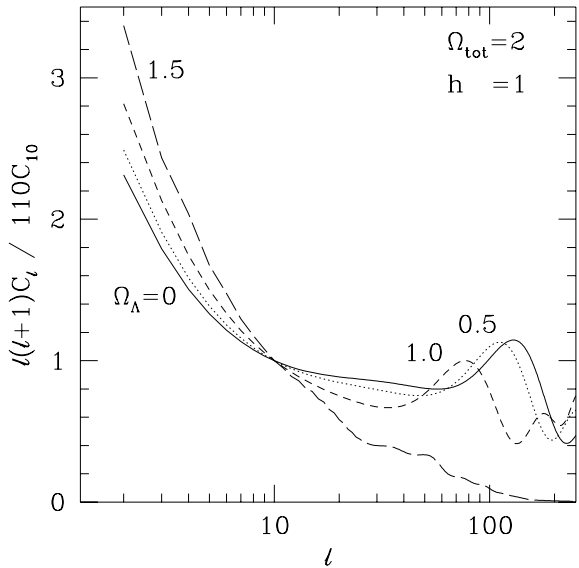
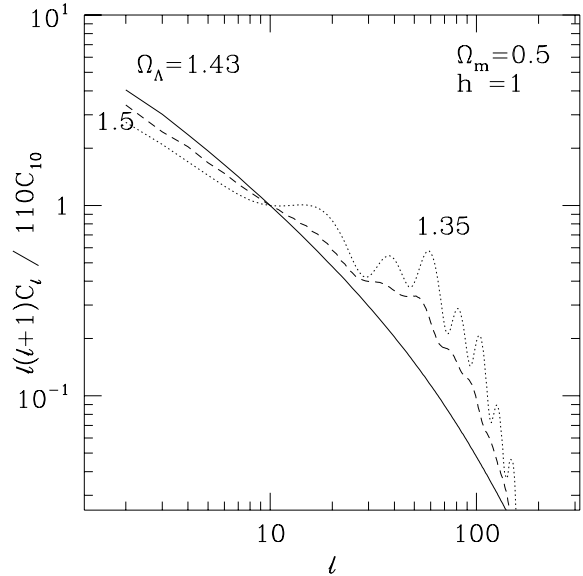

Figure 5: Angular power spectra for models with $\Omega_{tot} = 2$ and $\Omega_\Lambda \geq 0$. These models rise more steeply towards low $\ell$ than the $\Omega_\Lambda = 0$ curves, showing that for constraining $\Omega_{tot}$, the most conservative constraints are with $\Omega_\Lambda = 0$.

Figure 6: The solid line shows the angular power spectrum for a model with antipodal redshift at the last scattering surface: $z_a = z_{\text{rec}}$ (specifically: $\Omega_m = 0.5$; $\Omega_\Lambda = 1.43$). Note the absence of any structure in the spectrum. With $z_a = z_{\text{rec}}$ all structure is projected to $\ell = 0$, and we see only the damping tail plus some integrated Sachs-Wolfe effects at higher $\ell$. Also shown are two models where the peak is moved towards $\ell = 0$ ($\Omega_\Lambda = 1.35$; dotted) and where it has moved through $\ell = 0$ back to higher $\ell$ ($\Omega_\Lambda = 1.5$; dashed).

by the volume element, leading of course to finite fluctuations.

In the top right of Fig. 3 the peak location is $\ell_{\text{peak}} = 0$, which is just where the last scattering surface lies at the antipode. It had been conjectured that this would be a way to suppress CMB anisotropies (Durrer & Kovner 1990). We show in Fig. 6, the angular power spectrum taking into account the finite duration of last scattering, and the non-trivial gravitational effects on the spectrum, for a model with $\ell_{\text{peak}} = 0$ according to (18). Notice that the spectrum is featureless, showing only the "damping" tail of the anisotropies. However, it is non-zero because a finite width last scattering surface cannot project onto quite the whole sky, and furthermore there are still gravitational interactions on the photons. So we still see some anisotropies, but the acoustic peaks have all been magnified to ultra-large angular scales (very small $\ell$).

We also show on Fig. 6 two models where the last scattering epoch occurs slightly before and slightly after the antipodal redshift. Note that as we shift the antipode through $z = 1100$, the structure in the spectrum moves to $\ell \sim 0$ and back out to finite $\ell$ again. For all of these models the predicted large angle anisotropy (e.g. $C_2$) for fixed large-scale matter fluctuations is within a factor of 10 (in power, so $\sqrt{10}$ in temperature) of the standard cold dark matter result. Thus, though the power falls off rapidly with $\ell$, the suppression in power is not as extreme as one could naively imagine.

From Figs. 4–5 we see that on large scales (small $\ell$) the spectra are steeply falling functions of $\ell$. This was first noted by Abbott & Schaefer (1986), and is due to the decay of the gravitational potential at late times. A photon falling into a potential well which is decaying with time, has its in-fall blueshift only partly cancelled by its climb-out redshift. The net anisotropy builds up along the path, leading to larger temperature fluctuations on large scales (or long characteristic times where the effect has the most time to operate). This is known as the (late) integrated Sachs-Wolfe effect (Sachs & Wolfe 1967; see White et al. 1994, Hu, Sugiyama & Silk 1995, Hu & White 1995 for more discussion). Because this is the dominant effect at large scales, and does not appear in a universe with critical matter density, it makes scaling arguments difficult without accurate calculation of the anisotropies, evolved to the present epoch (c.f. Bunn, Scott & White 1995, White & Bunn 1995, Sugiyama 1995, Stompor, Gorski & Banday 1995).

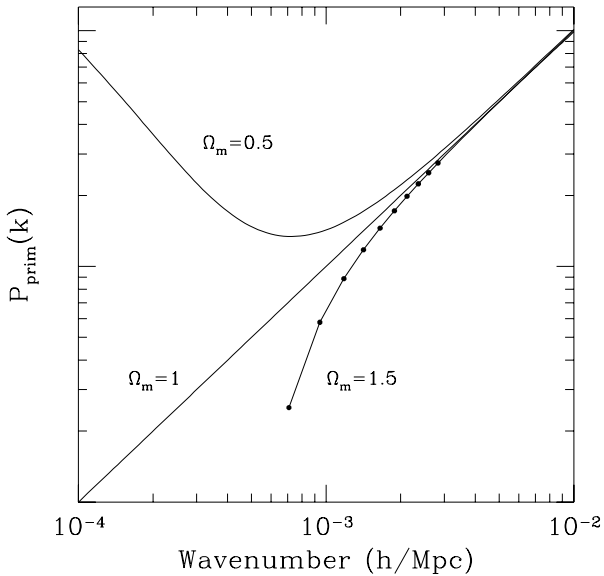

Figure 7: The primordial power spectrum for constant gravitational potential fluctuations, $\Phi$, per logarithmic interval in wavenumber, also called a "scale invariant" spectrum. The points on the $\Omega_m = 1.5$ curve indicate the quantized values of wavenumber, $\beta = 3, 4, 5\ldots$, in a closed universe. Note that only the $\Omega_m = 1$ model looks like a power law when plotted in this way, and that the behaviour for $\Omega_m > 1$ continues from the $\Omega_m < 1$ region as expected.

As can be seen from Fig. 3, the location of the first peak in the power spectrum can allow one to probe a large region of the $\Omega_m$–$\Omega_\Lambda$ plane. Data on anisotropies on degree scales are not yet up to the task of strongly constraining this position (Scott, Silk & White 1995), however we can still place new constraints on $\Omega_{\rm tot}$ from the large-angle fluctuations measured by the *COBE* satellite. We shall return to this in §7.

If however you believe in the existence of an acoustic peak in the anisotropy data, then this does place a strong constraint on closed models. It seems unlikely that the current data can be fit *without* some sort of peak at $\ell \sim 100$–200, unless several experiments have not been measuring primordial anisotropies. Taking the optimistic view that the data will indicate this more definitively in the near future, we may soon have a constraint that, for example, $\ell_{\rm peak} > 100$. From Fig. 3 we can see that this will rule out a substantial region of parameter space. We shall see in §7 that this region is probably already ruled out by COBE constraints alone. However, a tight constraint from the peak position would be insensitive to details of the primordial power spectrum around the curvature scale, where it is perhaps most dependent on specific inflationary models. Eventually we would expect an accurate measurement of the anisotropy spectrum to yield an extremely precise determination of $\Omega_{\rm tot}$.

## 6 Relation to Large Scale Structure

In Fig. 8 we show the relative normalization of the large-scale matter power spectrum and large-angle CMB angular power spectrum for a variety of models. Since much of the large angle CMB spectrum comes from the integrated Sachs-Wolfe effect, this normalization ratio cannot be obtained from simple growth factor arguments alone. We have calculated numerically the ratio of $B$ to $C_{10}$ where $C_{10}$ is as above, while

$$B \equiv \left. \frac{P(\beta)}{P_{\rm prim}(\beta)} \right|_{\beta=3} \qquad (21)$$

is the primordial power spectrum amplitude. We have defined this as the analogue of the quantity $B$ sometimes used to normalize the CDM power spectrum for $\Omega_m = 1$ (e.g. Bunn, Scott & White 1995). There $B$ is defined through $P(k) = B\, k\, T^2(k)$ (for an $n = 1$ spectrum) where $T(k)$ is the transfer function. Using $T(k \to 0) = 1$ we can write $B \equiv \lim_{k \to 0} P(k)/k$, from which (21) follows naturally, once you notice that a closed model has wavenumber $\beta\sqrt{K}$ quantized in units of the curvature scale (see appendix for details). Since the $\beta = 1$ and $\beta = 2$ modes are pure gauge modes (Abbott & Schaefer 1986) the smallest wavenumber (largest physical scale) is $\beta = 3$. We have chosen to use $C_{10}$, rather than a lower $\ell$, to minimize the primordial power spectrum dependence and reduce the contribution from the integrated Sachs-Wolfe effect. For $\ell = 10$ we are still very insensitive to details of cosmological processing and parameters other than $\Omega$.

For the matter power spectrum we also focus on the largest scales, to avoid complications due to the variable redshift of matter–radiation equality, which affects the amplitude of the fluctuations on smaller scales. As Fig. 8 shows, the matter–radiation normalization is a smooth function of $\Omega_{\rm tot}$ which depends upon the growth of fluctuations between $z = 1100$ and the present. For those cases we considered, which cover some extreme universes, the ratio of the matter to radiation power spectra varies by a factor of $\sim 10$ (cf. Sahni, Feldman & Stebbins 1992, and many earlier discussions of exponential growth during a loitering phase). The ratio for $\Omega_\Lambda = 0$ and $1 \leq \Omega_m \leq 2$ is well fit by

$$\frac{B}{C_{10}} = 320 \, \exp\left(-1.16 \, \Omega_m\right), \qquad (22)$$

in units of $(10^5 \, h^{-1} \, \mathrm{Mpc})^3$. The relation between the CMB and the *small*-scale matter fluctuations is given by Fig. 8 and the transfer function $T(k)$, to which we now turn.

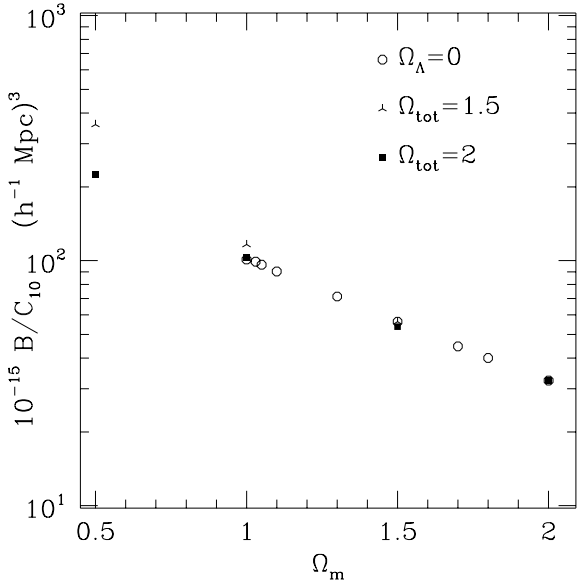
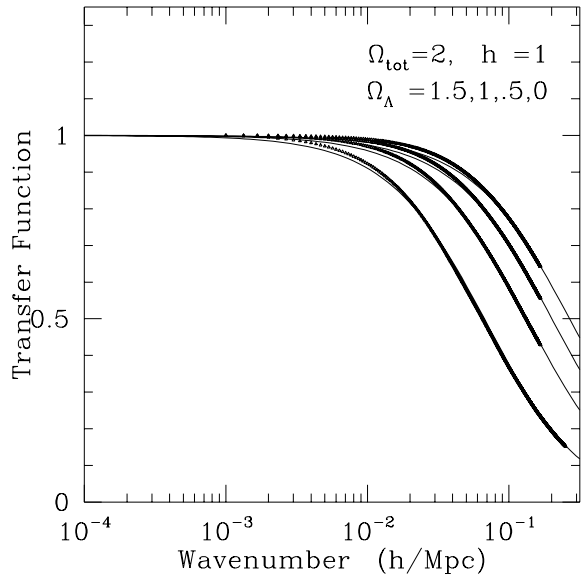

Figure 8: The relative normalization of the large-angle CMB and large-scale matter power spectra in closed models (see text for details). $B$ is a measure of the large-scale normalization of the matter power spectrum, while $C_{10}$ measures CMB fluctuations on scales smaller than the curvature scale but large enough to probe mostly potential fluctuations.

Figure 9: The matter transfer function. The solid lines are from the fitting function described in the text, while the symbols show the results of explicit calculations. Apart from a small discrepancy near the turn-over, which is caused mostly by the baryon content assumed for our models, the curves are a good fit to the numerical calculation.

For flat and open models it is known that the transfer function, $T(k)$, relating the initial and final power spectra is a function simply of the redshift of matter-radiation equality. Often one uses the fitting function (Efstathiou 1990)

$$T(k) = \left[1 + \left\{ak + (bk)^{3/2} + (ck)^2\right\}^\nu\right]^{-1/\nu}, \qquad (23)$$

with parameters $a = (6.4/\Gamma)$ Mpc, $b = (3.0/\Gamma)$ Mpc, $c = (1.7/\Gamma)$ Mpc, $\nu = 1.13$ and $\Gamma \simeq \Omega_m h$. Notice that perturbations on small scales, specifically those which entered the horizon when the universe was radiation dominated, have had their growth suppressed relative to those on larger scales.

We show in Fig. 9 that (23) is a good fit to the closed models as well. Thus, even though the growth rate can be "complicated" by the presence of loitering phases, $T(k)$ today is still fit by a 1 parameter family of curves. It is true that there are modes which can grow quickly during the loitering phase, when matter and cosmological constant conspire to keep the universe almost static as in the original analysis of (Jeans 1928). However, they will also have their growth retarded when the cosmological constant becomes dominant at later times and the universe begins to expand exponentially. These two effects approximately cancel, leading to fluctuation amplitudes vs. wavenumber that would be predicted from analogy with the open or flat models. [A similar conclusion for the large-scale (small $k$) modes was reached by Durrer & Kovner (1990).]

Using the large-scale normalizations of Fig. 8 and (23) one can calculate the moments of the power spec-trum for whatever parameters ($\Omega_m$, $\Omega_\Lambda$, $h$) are of interest. The "window functions" for moments such as $\sigma_8$, $V_{\rm rms}$, differ from those in a flat universe (see e.g. Wilson 1983) by geometric factors related to the 3-volume of the metric (i.e. $r \to \sin\chi$).

## 7 Limits on $\Omega$ from the CMB

There are already some weak constraints on flat $\Lambda$-dominated models from the shape of the large-angle power spectrum, amounting to $\Omega_\Lambda < 0.8$–$0.9$ (Bunn & Sugiyama 1995, Stompor, Gorski & Banday 1995, White & Bunn 1995). We can similarly use the 2-year *COBE* data to put a limit on $\Omega_{\rm tot}$ from the shape of the $C_\ell$'s at small $\ell$. This is quite insensitive to the cosmological parameters other than $\Omega_{\rm tot}$, making the limit reasonably robust.

First note from Fig. 5 that for fixed $\Omega_{\rm tot}$ the spectra with $\Omega_\Lambda > 0$ fall more steeply than with $\Omega_\Lambda = 0$. This is because there will always be *more* ISW effect in a $\Lambda$ model (Kofman & Starobinsky 1985), at redshifts where the cosmological constant starts to dominate: $z_\Lambda \sim \Omega_\Lambda(\Omega_{\rm tot} - 1)^{-1}$. Since the *COBE* data prefer spectra which are approximately "flat", or $C_\ell^{-1} \propto \ell(\ell+1)$, it follows that for fixed $\Omega_{\rm tot}$ models with $\Omega_\Lambda > 0$ will be less likely than the model with $\Omega_\Lambda = 0$. Also models with initial power spectra which have constant power in gravitational potential per logarithmic interval in wavenumber (scale invariant spectra) are "flatter" than models whose power spectra are proportional to wavenumber (see appendix and Fig. 7), and thus will be preferred by the *COBE* data.

In Fig. 10 we show the likelihood function vs. $\Omega_m$, integrated or marginalized over the normalization, for models with scale invariant spectra and $\Omega_\Lambda = 0$, fit to the *COBE* data. The likelihood function is normalized so that a "flat" spectrum, $C_\ell^{-1} \propto \ell(\ell+1)$, has $\mathcal{L} = 1$, and we have included a "prior" that $\Omega_m > 0.1$ (i.e. we set $\mathcal{L} = 0$ for $\Omega_m \leq 0.1$). The likelihood function for $\Omega_m < 1$ comes from (Górski et al. 1995) rescaled to the normalization of (White & Bunn 1995). The dashed line shows the cumulative likelihood, from which we estimate the 95% CL upper limit on $\Omega_m$ to be $\Omega_m \leq 1.5$. Any model which has $\Omega_{\rm tot} > 1.5$ will be less likely than the model with $\Omega_m = 1.5$ and $\Omega_\Lambda = 0$. If we take the upper limit from the cumulative likelihood to mean $\Omega_m = 1.5$ is statistically inadmissable at the 95% CL, we obtain the limit $\Omega_{\rm tot} \leq 1.5$.

This limit on $\Omega_{\rm tot}$ compares well with other limits on this quantity (Loh & Spillar 1986), but it is dependent on the initial power spectrum assumed. If the primordial power spectrum is "redder" than scale invariant (e.g. $n < 1$) then our limit is *strengthened*. However should a particular inflationary model predict $n > 1$ our limit would be weakened. A less model dependent limit could be placed on $\Omega_{\rm tot}$ from the position of the peaks in the angular power spectrum, when the data becomes less ambiguous.

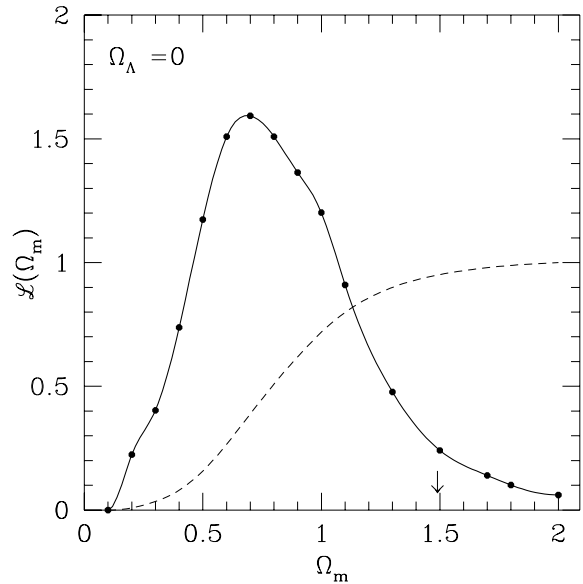

Figure 10: The likelihood function (solid) vs. $\Omega_m$, for models with scale invariant primordial spectra and $\Omega_\Lambda = 0$, fit to the *COBE* 2-year data and marginalized with respect to normalization. We have set $\mathcal{L}(\Omega_m) = 0$ for $\Omega_m \leq 0.1$ (our "prior"). The dashed line shows the cumulative likelihood, and the arrow indicates the 95% CL upper limit on $\Omega_m$. As discussed in the text this is a conservative upper limit on $\Omega_{\rm tot}$. The results for $\Omega_m < 1$ come from Gorski et al. (1995).

## 8 An Extreme Model

As an example of how non-standard models can sometimes be a surprisingly good fit to data, (Harrison 1993) considered a model with $\Omega_m = 10$ and $H_0 = 10\,{\rm km\,s^{-1}Mpc^{-1}}$. The idea was that the Hubble flow in such a universe could be so chaotic that our local estimates of $H_0$ could be far from the global value. With $h = 0.1$ the model is very old, while with $\Omega_m = 10$ it is very small and has almost finished its expansion and begun its (re)collapse. While we consider this model to be far from acceptable when examined in detail, we will use it here as an example of an *extreme* closed model. Harrison presents some scaling arguments for how structure formation and microwave background anisotropies would work out in such a model. Here we calculate $P(k)$ and the $C_\ell$ spectrum in detail.

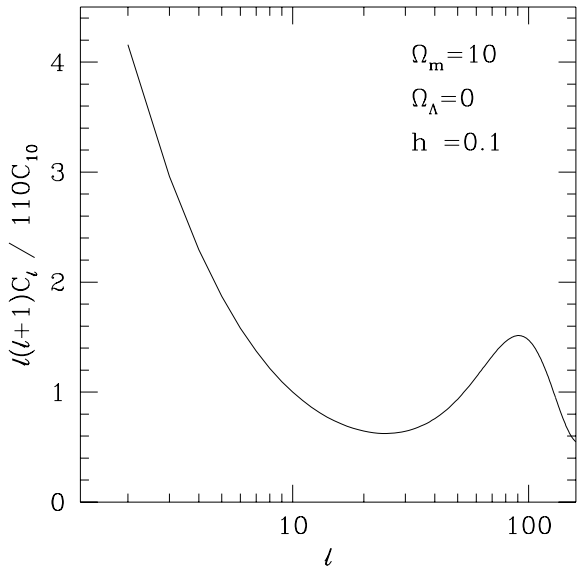

Figure 11: The angular power spectrum for the model of Harrison (1993) with $\Omega_m = 10$ and $h = 0.1$ assuming a scale invariant primordial spectrum and $\Omega_B h^2 = 0.0125$, $\Omega_\Lambda = 0$. Notice the peak in the power spectrum is at $\ell \simeq 100$, while the large-angle structure of the power spectrum is wildly inconsistent with the *COBE* data.

We shall assume that $\Omega_B = 1.25$ so that the model agrees roughly with Big Bang Nucleosynthesis determinations of $\Omega_B h^2$, though we can relax this assumption without changing our principle results. Since we know observationally that $\Delta T/T \sim 10^{-5}$ on large scales (Smoot et al. 1992) we can have confidence that our calculation of the angular power spectrum in linear perturbation theory will be accurate even in this extreme model.

The resulting anisotropy spectrum is presented in Fig. 11 for a scale invariant primordial power spectrum. One can see that the first peak occurs at $\ell \simeq 100$ (cf. Fig. 3; the height of the peak will depend on our assumption of the baryon content, but not its position) and at large angles it is wildly inconsistent with *COBE*. In linear theory the large-scale density fluctuation to large angle CMB fluctuation ratio is $B/C_{10} \sim 10 \ (10^5 \ h^{-1} \mathrm{Mpc})^3$. Thus, normalized to *COBE*, the large scale power in this model will be less than in e.g. standard CDM.

Once again the linear theory transfer function from explicit evolution of the density perturbations agrees well with (23). We find slightly less small scale power in the numerical calculation than for (23) with $\Gamma \simeq \Omega_m h = 1$, but only at the $\sim 10\%$ level for $k \leq 0.2 \ h/\mathrm{Mpc}$. A model with $\Gamma \simeq 1$ has much more small scale power than a standard CDM model, leading to earlier non-linearity at small scales. It would probably be in conflict with existing measurements of the large-scale matter power spectrum, which appear to require $\Gamma \sim 0.25$ (see e.g. Peacock & Dodds 1994). If we had decided to have $\Omega_B = \Omega_m = 10$ (in violation of nucleosynthesis bounds on $\Omega_B$) then the transfer function would damp much more strongly at high-$k$ than for the model with 90% dark matter, and would show oscillations similar to those which appear in the CMB anisotropy power spectrum.

## 9 Other Limits

The observational limits on $\Omega_\Lambda$ have been reviewed by (Carroll, Press & Turner 1992), so in this section we simply make a few additional remarks, referring the reader to that work for more details. Existing and future limits that are comparable to those from the CMB are discussed below.

Galaxy counts once seemed like a good way of measuring $\Omega_\Lambda$ through the change in the volume element (e.g. Loh & Spillar 1986), but separating the effects of geometry and evolution has proven to be the downfall of such methods (Gardner, Cowie & Wainscoat 1993). Nevertheless, the results from number counts have been interpreted as implying that $\Omega_m \leq 1.5$ (Loh & Spillar 1986), which is similar to our limit.

As pointed out in Gott, Park & Lee (1989), the existence of "normal looking" quasars at $z \geq 4$ puts a lower limit on the redshift of the antipode: $\sqrt{K} \left[ \eta_0 - \eta(z_a) \right] = \pi$. The contours of antipodal redshift are shown in Fig. 2, where we see that $z_a \geq 4$ rules out a *small* region in the top left of the plot. As discused in §5 the presence of structure in the CMB anisotropy spectrum would put a much stronger limit on the redshift of the antipode: $z_a \geq 1100$. The possible loop hole in this argument, namely that the last scattering surface lies beyond the antipode so that the structure "reappears" at $\ell \geq 0$, will be seen to be ruled out by gravitational lensing constraints. However we will see that these constraints are in any case stronger than the lower limits on $z_a$.

In a universe which becomes $\Omega_\Lambda$ dominated at some redshift, $z_\Lambda$, the expansion rate is dramatically differ-ent before and after $z_\Lambda$. Such a change can manifest itself as a "break" in the number of Ly$\alpha$ absorbers per unit redshift (Fukugita & Lahav 1991, Turner & Ikeuchi 1992). Future surveys of low-$z$ Ly$\alpha$ clouds may be able to put strong constraints on (or prefer) $\Omega_\Lambda$ models, but at present the situation is unclear.

Another sensitive probe of $\Omega_\Lambda$ in the future may come from separations of close quasar pairs (Phillipps 1994), as measured for example through the clustering properties of quasars in redshift surveys such as the Sloane Digital Sky Survey or the AAT 2dF. Here the idea is that the underlying correlation properties of quasars are expected to be isotropic, but when measured in terms of angular position and redshift one expects distortions (e.g. in the familiar $\pi - \sigma$ plane for various redshift shells). The comoving separation of a pair of objects at $z \pm \delta z$ and separated on the sky by $\delta\theta$ is (Phillipps 1994)

$$r^2 = g^2(\delta z)^2 + d_M^2(\delta\theta)^2, \qquad (24)$$

where $d_M(z)$ is the "proper motion" distance (Weinberg 1972) and

$$g(z) \equiv \left(1 - [\Omega_{\rm tot} - 1]H_0^2 d_M^2\right)^{-1/2} \frac{d\,d_M}{dz}. \qquad (25)$$

The ratio of $d_M$ to its derivative, $g$, at high redshift, which is a measure of the distortion of the isotropic correlation function, is strongly dependant on the cosmology assumed. With $10^5$ quasars this method could result in extremely strong constraints on (or a detection of) $\Lambda$, but awaits data from large surveys.

The strongest limits on $\Omega_\Lambda$ to date come from the rarity of gravitational lensing events (e.g. Fukugita, Futamase & Kasai 1990, Turner 1990, Fukugita et al. 1992, Kochanek 1993, Carroll, Press & Turner 1992). In a universe with a loitering phase the path length to any given redshift is dramatically enhanced, and the expected number of gravitational lenses along any line of sight (assuming a fixed density of objects) far exceeds the observed number. Define $P_{\rm lens}(z_s)$ as the probability of lensing a source at redshift $z_s$ relative to that in an $\Omega_m = 1$ universe (Carroll, Press & Turner 1992, Fukugita et al. 1992). The current surveys require $P_{\rm lens}(z_s \simeq 2) \leq 2\text{--}10$ depending on the detailed assumptions and data used; we shall take $P_{\rm lens} < 5$ to be a conservative limit. Contours of $P_{\rm lens}(z_s = 2)$ are shown in Fig. 2.

Since the lensing limit is currently more stringent than all of the other limits, the allowed region of parameter space becomes that to the lower right of some $P_{\rm lens} = 2\text{--}10$ line, and lower left of $\Omega_{\rm tot} = 1.5$. Less model dependent bounds on $\Omega_{\rm tot}$ could come from the location of the peak in the anisotropy spectrum, once the degree scale data becomes more robust. Taking $P_{\rm lens} \leq 5$ it is still possible to have $H_0 t_0 \simeq 1$ for a small region of parameter space with $\Omega_m \leq 0.5$. The lensing and CMB constraints are acting in different directions, thus an "improvement" in either will help to squeeze the allowed region in parameter space.

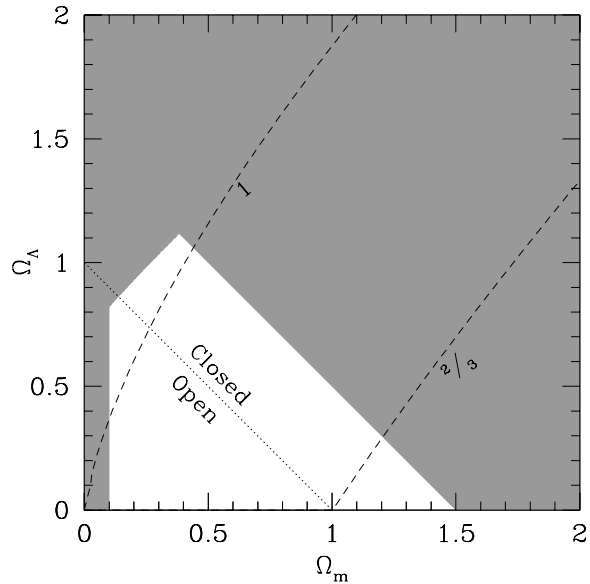

Figure 12: Constraints in the $\Omega_m$–$\Omega_\Lambda$ plane. The unshaded region is currently allowed by fairly conservative constraints. The three sides of the region come from lower limits on $\Omega_m$, upper limits on $P_{\rm lens}$ and an upper limit on $\Omega_{\rm tot}$ from large-angle CMB anisotropies. We also show two age lines with $H_0 t_0 = 2/3$ and $H_0 t_0 = 1$. Note that a fairly wide range of closed models have not yet been definitively ruled out, and that some of them are relatively old.

We show the currently allowed region of parameter space in Fig. 12, using dynamical, lensing and CMB constraints. We have chosen to be reasonably conservative in each case. The allowed region still contains a variety of closed models. We have indicated two representative age lines on this plot also. Note that for $H_0 = 80\,{\rm km\,s^{-1}Mpc^{-1}}$, the $H_0 t_0 = 1$ line has an age of 12.2 Gyr, while the $H_0 t_0 = 2/3$ line has an age of only 8.2 Gyr. Since dynamical estimates tend to give values like $\Omega_0 \sim 0.3$, it is perhaps tempting to speculate that the best bet for a model which is old enough, may be one with $\Omega_m + \Omega_\Lambda > 1$. However, it is also clear from the figure that the age gain over a $\Omega_{\rm tot} = 1$ model is not very great.

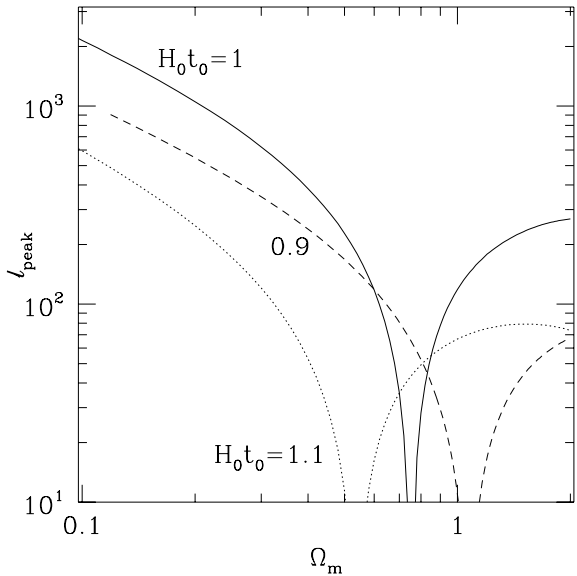

Figure 13: The position of the first peak in the CMB power spectrum, $\ell_{\rm peak}$, as a function of $\Omega_m$ for $H_0 t_0 = 1$ (solid), 1.1 (dotted), 0.9 (dashed). $\ell_{\rm peak}$ is a rapidly varying function of $\Omega_m$ along lines of constant $H_0 t_0$.

In the future we expect strong constraints on $H_0 t_0$ from a combination of large-scale structure (measuring $\Omega_m$) and the CMB. We show an example of this in Fig. 13, where we have plotted the position of the first peak in the CMB power spectrum as a function of $\Omega_m$, for three values of $H_0 t_0$. Note that the position of the peak in the experimentally favoured region is a rapidly varying function of $\Omega_m$.

In closing we note that large scale dynamics are mostly insensitive to $\Lambda$ (see e.g. Martel & Wasserman 1990, Martel 1991, Lahav et al. 1991). So, for example, the comparison of density and velocity fields measures the function $\Omega_m^{0.6}/b$ (where $b$ is the "bias" or amplitude of the matter power spectrum) independently of whether the model contains a contribution from $\Lambda$ which makes it flat. In the limit that the dependence of the perturbation growth rate is a weak function of $\Omega_\Lambda$, these conclusions apply equally well in closed models. Dynamical tests thus generally tend to measure $\Omega_m$ rather than, say, $\Omega_{\rm tot}$. In marked contrast the location of the peaks in the CMB anisotropy angular power spectrum depends essentially only on the geometry, and thus measures both $\Omega_m$ and $\Omega_\Lambda$, largely through the quantity $\Omega_{\rm tot}$. This makes the combination of the two an extremely powerful constraint on cosmological models.

## 10 Conclusions

We have examined CMB anisotropies and large-scale structure formation in models with positive spatial curvature, i.e. closed models. We presented for the first time the angular power spectrum of the anisotropies over a range of angular scales. The relative normalization of the CMB anisotropies and the large-scale matter power spectrum show behaviour which is a simple continuation from the flat case.

For scale-invariant potential fluctuations, the anisotropies are steeply rising towards low multipoles. With $\Lambda = 0$ the position of the main acoustic peak scales approximately as $\ell_{\rm peak} \propto \Omega_{\rm tot}^{-1/2}$ as expected. There is a richer structure (shown in Fig. 3) when $\Lambda \neq 0$. It is possible in principle to have last-scattering occur at the antipode, in which case the spectrum looks very different. However, this case is unlikely to occur for our observed universe.

As a particular example we have calculated the anisotropy spectrum for the model postulated by Harrison (1993), which has $\Omega_m = 10$ and $h = 0.1$. Since the CMB fluctuations are measured to be well in the linear regime, the chaotic behaviour envisioned by Harrison to account for our "misinterpretation" of the current data is hardly relevant, so one can say with confidence that the model is in conflict with current CMB and large-scale structure data.

Extending the likelihood function vs. $\Omega$ from the COBE 2-year data to $\Omega \geq 1$ we obtain a 95% CL upper limit of $\Omega_{\rm tot} \leq 1.5$ for primordial spectra with $n \leq 1$, which compares favorably with existing limits on $\Omega$. This limit is dependent on our assumption of a primordial power spectrum and the absence of tensor modes (which are self-consistent), however a more secure bound could come from the position of the first peak in the angular power spectrum as shown in Fig. 3.

The low number of gravitationally lensed quasars is currently the tightest constraint on the value of $\Omega_\Lambda$. However, limits from large-angle CMB anisotropies, as well as the position of the acoustic peak, show great promise for the near future. It is also important to note that these effects can constrain different combinations of $\Omega_m$ and $\Omega_\Lambda$ than lensing and large-scale structure (which measures largely $\Omega_m$). If we were able to measure $\Omega_m$ accurately (from the matter power spectrum for example), then for $\Omega_m < 0.5$ an "old" universe has a strong dependence of $\ell_{\rm peak}$ on $\Omega_m$ (see Fig. 13). Thus it should be easy to measure either $\Omega_\Lambda$ or $H_0 t_0$ from the peak location in such models, given the other. This is just one consequence of the age contour lines running roughly perpendicular to $\ell_{\text{peak}}$ contour lines in the $(\Omega_m, \Omega_\Lambda)$ plane.

There is still a small region of parameter space which is allowed by cosmological constraints, and which lies in the $\Omega_{\text{tot}} > 1$ part which is oftenneglected (Fig. 12). For some smaller part of this region the age of the universe is also favourable for the high values of the Hubble constant which seem most commonly to have been discussed in the last few years. With this motivation, but largely because they are still viable today, we believe these closed models are worthy of examination.

We wish to thank Wayne Hu for many useful conversations and Ted Bunn for help with the *COBE* data.

## A  Perturbations in Non-Flat Geometries

In this appendix we provide a derivation of the Einstein, fluid and Boltzmann equations in a Friedmann-Robertson-Walker universe with arbitrary spatial curvature, paying particular attention to the case of a closed universe (spherical geometry). Our derivation is based upon the pioneering paper of Wilson (1983). We use his formalism, but extended and with a more numerically stable normalization scheme (from Abbott & Schaefer 1986, who presented the first calculation of the large angle anisotropies in a closed universe). This normalization is also used by Hu & Sugiyama (1995). We have also drawn on the work of Gouda, Sugiyama & Sasaki (1991) which provides some (but not all) of the steps missing from both Wilson (1983) and Wilson's (1981) Ph.D. thesis. Our intention is to provide a complete derivation of the equations which are needed to evolve cosmological perturbations in a curved background.

In order to keep this appendix to a manageable size, we assume that the reader is familiar with the derivation and implimentation of the fluid and Boltzmann equations in a flat universe. An excellent reference for this case is Ma & Bertschinger (1995), and our notation is based partly upon that work. Further details can be found in HSSW and references therein. We will only concern ourselves with scalar perturbations, although following Abbott & Schaefer (1986) it would be relatively straightforward to generalize to vectors or tensors. An analytic treatment of the anisotropies induced by tensors has been presented by Allen, Caldwell & Koranda (1995).

### A.1  The Metric

We define our metric to be

$$ds^2 = -dt^2 + a^2(t)\left(\gamma_{ij} + h_{ij}\right) dx^i dx^j \tag{26}$$

$$= a^2(t)\left[-d\eta^2 + \left(\gamma_{ij} + h_{ij}\right) dx^i dx^j\right], \tag{27}$$

where $\gamma_{ij}$ is a 3-metric with constant curvature $K = H_0^2(\Omega_{\rm tot}-1)$, $\eta \equiv \int dt/a(t)$ is the conformal time and $h_{ij}$ is the metric perturbation in synchronous gauge (Lifshitz 1946, Landau & Lifshitz 1975,§97,Weinberg 1972, Peebles 1980, Peebles 1993). In our notation latin indices run over spatial components from $1,2,3$ and greek indices over space and (conformal) time components $0,1,2,3$.

By definition, the geodesic equations are those which extremize the Lagrangian, $\mathcal{L} = \tfrac{1}{2} g_{\mu\nu}\dot{x}^\mu \dot{x}^\nu$ (where an overdot here *only* represents a derivative with respect to some parameter, $\lambda$, along the path). Equating the Euler-Lagrange equations for $\mathcal{L}$ with the other form of the geodesic equation,

$$\frac{d^2 x^\mu}{d\lambda^2} + \Gamma^\mu_{\nu\rho}\frac{dx^\nu}{d\lambda}\frac{dx^\rho}{d\lambda} = 0 \ , \tag{28}$$

we can read off the connection coefficients:

$$\begin{aligned}
\Gamma^0_{00} &= (\dot{a}/a) \\
\Gamma^0_{ij} &= (\dot{a}/a)(\gamma_{ij} + h_{ij}) + \tfrac{1}{2}\dot{h}_{ij} \\
\Gamma^j_{0i} &= (\dot{a}/a)\delta^j_i + \tfrac{1}{2}\dot{h}^j_i \\
\Gamma^j_{ik} &= {}^{(3)}\Gamma^j_{ik} + \tfrac{1}{2}\left((h^j_i)_{|k} + (h^j_k)_{|i} + (h_{ik})^{|j}\right) .
\end{aligned} \tag{29}$$

Here ${}^{(3)}\Gamma$ is the connection for the spatial metric $\gamma_{ij}$ of constant curvature, and a subscript $|i$ means a covariant derivative wth respect to this metric. This is distinguished from $;\alpha$ which refers to a covariant derivative with respect to the full 4-dimensional metric. An overdot means $d/d\eta$ henceforth.

All that we will need explicitly from $\gamma_{ij}$ is the curvature tensor

$${}^{(3)}R^m_{ijk} = K\left(\delta^m_j \gamma_{ik} - \delta^m_k \gamma_{ij}\right), \tag{30}$$

which allows us to commute covariant derivatives, e.g. in the simplest case $V_{i|jk} - V_{i|kj} = {}^{(3)}R^m_{ijk}V_m$. However, we shall be concerned with commuting pairs of indices among arbitrarily many:

$$\begin{aligned}
V_{|i_1 i_2 \cdots i_\ell ij \cdots k} &= \left(V_{|i_1 i_2 \cdots i_\ell i}\right)_{|j \cdots k} \\
&= \left(V_{|i_1 i_2 \cdots i i_\ell} + \sum_{n=1}^{\ell-1} {}^{(3)}R^{j_n}_{i_n i_\ell i} V_{|i_1 \cdots j_n \cdots i_{\ell-1}}\right)_{|j \cdots k} .
\end{aligned} \tag{31}$$

Note that indices *after* the pair to be commuted go along for the ride.

For deriving the field equations we will also need the Ricci tensor:

$$R_{\alpha\beta} = (\Gamma^{\gamma}_{\alpha\beta})_{,\gamma} - (\Gamma^{\gamma}_{\alpha\gamma})_{,\beta} + \Gamma^{\gamma}_{\sigma\gamma}\Gamma^{\sigma}_{\alpha\beta} - \Gamma^{\gamma}_{\beta\sigma}\Gamma^{\sigma}_{\alpha\gamma} \tag{32}$$

## A.2 Basis Functions

In linear theory, modes with different $\vec{k}$ evolve independently, so working in a wavenumber basis is advantageous. Although in a non-flat universe the Fourier transform is not easily defined, we can still work in terms of the eigenfunctions of the Laplacian operator, which we shall call $Q(\vec{x})$. This is just $e^{i\vec{k}\cdot\vec{x}}$ in flat space; a detailed discussion of representations and spectra of $Q(x)$ can be found in Abbott & Schaefer (1986). We will only need the defining relation:

$$\begin{aligned}\nabla^2 Q \equiv \gamma^{ij}Q_{|ij} &= -k^2 Q \\ &= -(\beta^2 - 1)KQ.\end{aligned} \tag{33}$$

The last equality defines $\beta$ for $\Omega_{\text{tot}} > 1$. In this case the spectrum of these functions are the integers $\beta = 3, 4, 5\ldots$, since the spatial hypersurfaces have a spherical geometry (cf. Fourier series on a circle). The modes $\beta = 1, 2$ correspond to pure gauge modes (Abbott & Schaefer 1986), and are not included in the spectrum.

We can expand our perturbed quantities in terms of these functions and their covariant derivatives. For example the metric perturbation can be written

$$h_{ij}(\vec{x}, \eta) = \frac{1}{3}\left[h + H\right]Q\gamma_{ij} + k^{-2}HQ_{|ij}, \tag{34}$$

where $h$ is the trace of the perturbation and $H$ is the traceless part.

## A.3 Einstein and Fluid Equations

The evolution equations for the background metric and its perturbation follow from Einstein's equations, which can be written

$$R_{\alpha\beta} = 8\pi G\left(T_{\alpha\beta} - \frac{1}{2}g_{\alpha\beta}T\right) \equiv 8\pi G\widetilde{T}_{\alpha\beta}, \tag{35}$$

where $T_{\alpha\beta}$ is the stress-energy tensor and $T$ is its trace. The background stress-energy tensor is of the perfect fluid form (with density $\bar{\rho}$ and pressure $\bar{p}$) plus a perturbation:

$$\begin{aligned}T^0_0 &= -\bar{\rho}\left(1 + Q\delta\right), \tag{36}\\ T^j_0 &= -(\bar{\rho} + \bar{p})\,v(ik)^{-1}Q^{|j}, \tag{37}\\ T^i_j &= (\bar{p} + Q[\delta p])\,\delta^i_j + \Sigma^i_j, \tag{38}\end{aligned}$$

with $\Sigma^i_j$ the anisotropic stress ($\Sigma^i_i = 0$), which we will not need explicitly.

The required evolution equations come from the time-time and time-space Einstein's equations. The unperturbed equations are the familiar Friedmann-Robertson-Walker equations for the evolution of the scale factor $a(\eta)$. To first order in the perturbation (first-order quantities denoted by the pre-superscript (1)):

$$^{(1)}R_{00} = -\frac{1}{2}\left(\ddot{h} + \frac{\dot{a}}{a}\dot{h}\right)Q \tag{39}$$

$$\text{and} \quad ^{(1)}\widetilde{T}_{00} = \frac{1}{2}a^2\left(\bar{\rho}\,\delta + 3[\delta p]\right)Q, \tag{40}$$

so that

$$\frac{d}{d\eta}\left(a\dot{h}\right) = 8\pi G a^3\,\bar{\rho}\,(1 + 3w)\,\delta, \tag{41}$$

with $w = \bar{p}/\bar{\rho}$. The time-space equation involves

$$\begin{aligned}2 \times {}^{(1)}R_{0i} &= \left(\dot{h}^i_j\right)_{|j} - \text{Tr}\left(\dot{h}\right)_{|i} \\ &= -\dot{h}Q_{|i} + \left[\frac{1}{3}\left(\dot{h} + \dot{H}\right)Q_{|i} + k^{-2}\dot{H}(Q^{|j}_{|i})_{|j}\right] \\ &= -\frac{2}{3}\left(\dot{h} + \dot{H}k^{-2}(k^2 - 3K)\right)Q_{|i}.\end{aligned} \tag{42}$$

In the last line we have made use of the following identity (coming from equation 31):

$$\begin{aligned}
\gamma^{jk} Q_{|ijk} &= \gamma^{jk} Q_{|jik} \\
&= \gamma^{jk} \left( Q_{|jki} + {}^{(3)}R^m_{jik} Q_{|m} \right) \\
&= -k^2 Q_{|i} + \gamma^{jk} K \left( \delta^m_i \gamma_{jk} - \delta^m_k \gamma_{ji} \right) Q_{|m} \\
&= \left( -k^2 + 2K \right) Q_{|i}.
\end{aligned} \qquad (43)$$

The stress energy tensor is

$$^{(1)}\widetilde{T}_{0i} = -a^2 \left( \bar{\rho} + \bar{p} \right) v \left( ik \right)^{-1} Q_{|i}. \qquad (44)$$

If we redefine $H$ to absorb the factor $(1 - 3K/k^2)$ in (42), then the evolution equation is

$$\frac{k^2}{3} \left( \dot{h} + \dot{H} \right) = -8\pi G a^2 \bar{\rho} (1 + w) \theta, \qquad (45)$$

where $\theta = i\vec{k} \cdot \vec{v}$.

The equations for the non-relativistic fluids (CDM and baryons) can be simply obtained from the conservation equation for the stress-energy tensor: $T^\beta_{\alpha;\beta} = 0$, or

$$\frac{1}{\sqrt{-g}} \left( \sqrt{-g} T^\beta_\alpha \right)_{,\beta} = \frac{1}{2} g_{\beta\gamma,\alpha} T^{\beta\gamma}. \qquad (46)$$

We will show later that for fluids without anisotropic stress the fluid equations are unchanged from their flat space counterparts. Explicit evaluation of these equations in the $K = 0$, $\bar{p} = 0$ case is straightforward. Using $\sqrt{-g} = a^4(1 + \frac{1}{2}h)$, we find for the $\alpha = 0$ equation

$$\dot{\delta} = -\left( \theta + \frac{1}{2}\dot{h} \right), \qquad (47)$$

while with $\alpha = i$ we obtain

$$\dot{\theta} = -\frac{\dot{a}}{a} \theta. \qquad (48)$$

### A.4 Legendre Tensors

The radiation distribution function will exhibit an azimuthal symmetry, so we would like to use an expanion in Legendre polynomials. To avoid introducing a fixed basis, it is convenient to define the following functions of position $\vec{x}$ and an angle $\hat{n}$,

$$\begin{aligned}
P_{(0)} &= 1, \\
P^i_{(1)} &= n^i, \\
P^{ij}_{(2)} &= \frac{1}{2} \left( 3n^i n^j - \gamma^{ij} \right) \\
P^{ijk}_{(3)} &= \frac{1}{2} \left( 5n^i n^j n^k - 3\gamma^{(ij} n^{k)} \right), \\
(2\ell + 1) n^{(i_1} P^{i_2 \cdots i_\ell i_{\ell+1})}_{(\ell)} &= \ell \gamma^{(i_1 i_2} P^{i_3 \cdots i_{\ell+1})}_{(\ell-1)} + (\ell + 1) P^{i_1 \cdots i_{\ell+1}}_{(\ell+1)},
\end{aligned} \qquad (49)$$

where parentheses denote symmetrization in the indices. Each $P_{(\ell)}$ is a function of position through $\gamma_{ij}$ only (so its covariant derivative vanishes), and is explicitly symmetric. As can be seen from the definition, $P_{(\ell)}$ is a "Legendre polynomial with dangling indices", i.e. rather than a polynomial of a dot-product of vectors, the tensor has a free index for each power of the vector argument. Later this will allow us to perform an expansion in Legendre polynomials by contracting tensors in an explicitly coordinate free way. Clearly

$$\hat{k}_{i_1} \ldots \hat{k}_{i_\ell} P^{i_1 \cdots i_\ell}_{(\ell)} = P_\ell(\hat{k} \cdot \hat{n}) \qquad (50)$$

for any unit vector $\hat{k}$.

Some properties of $P_{(\ell)}$ which will be useful are

$$\gamma_{ij}P^{ij\cdots k}_{(\ell)} = 0, \tag{51}$$

$$n_i P^{ij\cdots k}_{(\ell+1)} = P^{j\cdots k}_{(\ell)}, \tag{52}$$

which are now proven by induction. Both are true for $\ell = 0, 1, 2, 3$ by inspection.

Assuming (52) holds true for all $P_{(i)}$ with $i < \ell + 1$, we have

$$
\begin{aligned}
n_i \gamma^{(ii_1} P^{i_2\cdots i_\ell)}_{(\ell-1)} &= \frac{2}{\ell(\ell+1)} \left[ \frac{\ell(\ell-1)}{2} \gamma^{(i_1 i_2} P^{i_3\cdots i_\ell)}_{(\ell-2)} + \ell\, n^{(i_1} P^{i_2\cdots i_\ell)}_{(\ell-1)} \right] \\
&= \frac{1}{\ell+1} \left[ \left\{ (2\ell - 1) n^{(i_1} P^{i_2\cdots i_\ell)}_{(\ell-1)} - \ell P^{i_1\cdots i_\ell}_{(\ell)} \right\} + 2 n^{(i_1} P^{i_2\cdots i_\ell)}_{(\ell-1)} \right] \\
&= \frac{1}{\ell+1} \left[ (2\ell + 1) n^{(i_1} P^{i_2\cdots i_\ell)}_{(\ell-1)} - \ell P^{i_1\cdots i_\ell}_{(\ell)} \right]
\end{aligned}
\tag{53}
$$

where the factors $\ell(\ell-1)/2$ and $\ell$ are the number of terms in the symmetric products. Also

$$(2\ell + 1) n_i n^{(i} P^{i_1\cdots i_\ell)}_{(\ell)} = \frac{2\ell+1}{\ell+1} \left[ P^{i_1\cdots i_\ell}_{(\ell)} + \ell n^{(i_1} P^{i_2\cdots i_\ell)}_{(\ell-1)} \right] \quad . \tag{54}$$

Putting these together gives equation (52).

Similarly assuming (51) holds true for all $P_{(i)}$ with $i < \ell + 1$, we have

$$
\begin{aligned}
(\ell+1) \gamma_{i_1 i_2} P^{i_1\cdots i_{\ell+1}}_{(\ell+1)} &= (2\ell+1) \gamma_{i_1 i_2} n^{(i_1} P^{i_2\cdots i_{\ell+1})}_{(\ell)} - \ell \gamma_{i_1 i_2} \gamma^{(i_1 i_2} P^{i_3\cdots i_{\ell+1})}_{(\ell-1)} \\
&= \frac{(2\ell+1)}{(\ell+1)} 2 P^{i_3\cdots i_{\ell+1}}_{(\ell-1)} - \frac{2}{\ell(\ell+1)} \ell \left( 3 P^{i_3\cdots i_{\ell+1}}_{(\ell-1)} + 2(\ell-1) P^{i_3\cdots i_{\ell+1}}_{(\ell-1)} \right) \\
&= \frac{1}{\ell+1} P^{i_3\cdots i_{\ell+1}}_{(\ell-1)} \left[ 2(2\ell+1) - 2(2\ell+1) \right] \\
&= 0,
\end{aligned}
\tag{55}
$$

where the factors of $\ell+1$ and $\ell(\ell+1)/2$ are the number of terms in the symmetric products.

### A.5  Recurrence Relation

Later we will expand our radiation distribution function in terms of mode functions

$$G_\ell \equiv (-k)^{-\ell}\, Q_{|i_1\cdots i_\ell}(\vec{x};\vec{k})\, P^{i_1\cdots i_\ell}_{(\ell)}(\vec{x},\hat{n}). \tag{56}$$

The interesting property of these functions for our purpose is the recurrence relation that they satisfy,

$$n^i G_{\ell|i} = k \left[ \frac{\ell}{2\ell+1} \kappa_\ell^2 G_{\ell-1} - \frac{\ell+1}{2\ell+1} G_{\ell+1} \right], \tag{57}$$

where

$$
\begin{aligned}
\kappa_0^2 &\equiv 1 \\
\kappa_\ell^2 &\equiv 1 - (\ell^2 - 1) K/k^2 \qquad \ell \geq 1.
\end{aligned}
\tag{58}
$$

Notice that in terms of $\beta$ we can write $\kappa_\ell = \sqrt{(\beta^2 - \ell^2)/(\beta^2 - 1)}$ for $\ell \geq 1$.

The proof of this relation is simply an exercise in combinatorics and commuting covariant derivatives using (31). Let us first prove some identities which will be crucial. The simplest is

$$Q_{|i_1\cdots i_\ell i}\, n^i P^{i_1\cdots i_\ell}_{(\ell)} = Q_{|i_1\cdots i_\ell i}\, n^{(i} P^{i_1\cdots i_\ell)}_{(\ell)} + \frac{1}{3}\ell(\ell-1) K (-k)^{\ell-1} G_{\ell-1}\, , \tag{59}$$

which comes from commuting the $i$ to all $\ell+1$ possible positions (recall that $P_{(\ell)}$ is symmetric in all indices). An index swap of $i$ in the $(n+1)$th place with $i_n$ in the $n$th place gives $(n-1)$ factors of $KG_{\ell-1}$. The number of factors obtained when symmetrizing $n^i P_{(\ell)}$ is thus

$$\sum_{j=0}^{\ell-1} j(j+1) = \frac{1}{3}\ell(\ell^2 - 1) \; , \tag{60}$$

which when divided by the $\ell+1$ terms gives the relation above.

A more taxing but just as conceptually simple identity is

$$Q_{|i_1\cdots i_\ell i} \gamma^{(i i_1} P_{(\ell-1)}^{i_2\cdots i_\ell)} = \frac{2}{\ell(\ell+1)}\left(-k^2 \frac{\ell(\ell+1)}{2} + \frac{(\ell+2)!}{6(\ell-2)!}K\right)(-k)^{\ell-1}G_{\ell-1} \; , \tag{61}$$

which comes from looking at contractions of the two indices, $i$ and $i_\ell$, in arbitrary positions on $Q_|$ with $\gamma^{i i_\ell} P_{(\ell-1)}^{i_1\cdots i_{\ell-1}}$. If the indices are in the first two positions then one obtains $(-k^2 Q)_{|i_1\cdots i_{\ell-1}}$ from the eigenvalue equation for $Q$. Thus the problem is: how many factors of $K$ does commuting indices introduce before these two indices are in the first two positions? If we decide that $i_\ell$ will always come before $i$ (otherwise use the symmetry of $\gamma^{i i_\ell}$ and relabel these indices) then we first shift $i_\ell$ from the $n$th position to the first, then shift $i$ from the $m$th position to the second position. Moving something from the $n$th position to the $(n-1)$th position, assuming that it is the first index on $\gamma^{i i_\ell}$ gives $(n-2)$ factors of K. Moving this index all the way to the left thus introduces $(n-1)(n-2)/2$ factors of $K$. The second index is more complicated since now the possible contractions of indices allow $\gamma_{ij}\gamma^{ij} = 3$ and $\gamma_{ij}\gamma^{jk} = \delta_i^k$. One finds that moving the $m$th index to the second position gives $m(m-1)/2 - 1$ factors of $K$. The total number of $K$'s is thus

$$\sum_{n=1}^{\ell} \sum_{m=n+1}^{\ell+1} \frac{(n-1)(n-2)}{2} + \left[\frac{m(m-1)}{2} - 1\right] = \frac{(\ell+2)!}{6(\ell-2)!} \; . \tag{62}$$

Now we are in a position to derive the recurrence relation:

$$\begin{aligned}
n^i G_{\ell|i} &= (-k)^{-\ell}\left[Q_{|i_1\cdots i_\ell} n^i P_{(\ell)}^{i_1\cdots i_\ell}\right] \\
&= (-k)^{-\ell}\left[Q_{|i_1\cdots i_\ell} n^{(i} P_{(\ell)}^{i_1\cdots i_\ell)} + \frac{1}{3}\ell(\ell-1)K(-k)^{\ell-1}G_{\ell-1}\right] \\
&= (-k)^{-\ell}\left[\frac{1}{3}\ell(\ell-1)K(-k)^{\ell-1}G_{\ell-1} + Q_{|i_1\cdots i_\ell i}\left(\frac{\ell}{2\ell+1}\gamma^{(i i_\ell} P_{(\ell-1)}^{i_1\cdots i_{\ell-1})} + \frac{\ell+1}{2\ell+1}P_{(\ell+1)}^{i_1\cdots i_\ell i}\right)\right] \\
&= -k\frac{\ell+1}{2\ell+1}G_{\ell+1} + (-k)^{-\ell}\left[\frac{1}{3}\ell(\ell-1)K(-k)^{\ell-1}G_{\ell-1} + \right. \\
&\quad \left. \frac{\ell}{2\ell+1}\frac{2}{\ell(\ell+1)}\left(\frac{\ell(\ell+1)}{2}(-k^2) + \frac{(\ell+2)!}{6(\ell-2)!}K\right)(-k)^{\ell-1}G_{\ell-1}\right] \\
&= k\left[\frac{\ell}{2\ell+1}\kappa_\ell^2 G_{\ell-1} - \frac{\ell+1}{2\ell+1}G_{\ell+1}\right] \; . \tag{63}
\end{aligned}$$

Let us try to understand this relation physically. As we shall see in the next section this recurrence relation describes the way power is transferred between $\ell$ modes by propagation. Consider first the case $\Omega_{\rm tot} = 1$, or $K = 0$. A given physical scale subtends a smaller and smaller angle (higher $\ell$) as the distance between emission and reception of the photons increases. This is characterized by the coupling of the $\ell$ modes in the recurrence relation, which allow power to be transferred from low $\ell$ to high $\ell$. In the case of a non-flat universe ($K \neq 0$), the manner in which power is transferred from low to high $\ell$ is modified by the effects of curvature (geodesics are no longer straight lines). For an open universe ($K < 0$) the coupling between modes is enhanced, making it easier to transfer power to higher $\ell$ (smaller scales). This is because geodesics in an open universe diverge, thus a fixed physical scale subtends a smaller angular scale at the same distance in an open universe. In a closed universe ($K > 0$) the coupling between modes is decreased, making it harder to move power to higher $\ell$. In fact when $k^2/K = \ell^2 - 1$, $\kappa_\ell = 0$ and there is no coupling to higher modes! In a closed universe *there is a finite heirarchy for each mode*. This reflects the fact that in a closed universe, there is a maximum physical scale that is probed by geodesics with fixed angular separation. Note also that once power reaches this $\ell$ it will reflect back to lower $\ell$, showing that the same physical scale begins to subtend a larger angle after the maximum separation of the geodesics has been reached (at the "equator"). It is this property which is reponsible for the behaviour of the position of the first peak in the power spectrum (see Fig. 3).

One last important property of the $G_\ell$ is their normalization. Notice that $Q(\vec{x}; \vec{k})$ depends on the direction of $\vec{k}$. Recalling that $P_{(\ell)}$ is a symmetric tensor and (50, 51), we have that

$$\int d\Omega_k \; G_\ell \propto \int d\Omega_k \; QP_\ell(\hat{k} \cdot \hat{n}). \tag{64}$$

To fix the normalization of these functions we use the recurrence relation and

$$\int d\Omega_k \int \sqrt{-g} \, d^3x \; n^i \, [G_\ell G_{\ell-1}]_{|i} = 0, \tag{65}$$

to find

$$\int d\Omega_k \; ||G_\ell^2|| = \kappa_\ell^2 \int d\Omega_k \; ||G_{\ell-1}^2||. \tag{66}$$

### A.6 The Boltzmann Equation

We are now in a position to derive the Boltzmann equation for the radiation distribution function, which is a function of the 6 dimensional phase space for the radiation $(\vec{x}, \vec{p})$ and the conformal time $\eta$. We scale out the redshifting of the photon momentum, $\vec{p}$, by the definition $\vec{q} = a\vec{p}$ and write $\vec{q} = q\hat{n}$. Integrating $f(\vec{x}, \hat{n}, q, \eta)$ over $q$ gives the brightness perturbation, $F(\vec{x}, \hat{n}, \eta)$, which we can expand in terms of our basis functions as

$$F(\vec{x}, \hat{n}, \eta) = \sum_{\vec{k}} \sum_\ell F_\ell(\vec{k}, \eta) \left( \prod_\ell \kappa_\ell \right)^{-1} G_\ell(\vec{x}, \hat{n}), \tag{67}$$

where the factors of $\kappa_\ell$ have been included to provide a convenient normalization of the functions (see equation 66). By including the factors of $\kappa_\ell$ we ensure that the expression for $C_\ell$ in terms of $F_\ell$ will be formally unchanged from the $K = 0$ case. Absorbing the very large factorials involved in $\prod_\ell \kappa_\ell$ is also important to ensure numerical stability in the computation of the radiation power spectrum.

The Boltzmann equation for the radiation can be written schematically as

$$\begin{aligned} \frac{DF}{D\eta} &\equiv \frac{\partial F}{\partial \eta} + \frac{d}{d\eta} F\left(\vec{x}(\eta), \hat{n}\right) + \dot{q} \frac{\partial F}{\partial q} \\ &\approx \frac{\partial F}{\partial \eta} + \dot{x}^i \frac{\partial F}{\partial x^i} + \dot{n}^i \frac{\partial F}{\partial n^i} + \dot{q} \frac{\partial F}{\partial q} \\ &= \text{Collision Term}. \end{aligned} \tag{68}$$

For the photons the RHS is the collision integral for Thomson scattering, which can be found in Hu, Scott & Silk (1994) and Dodelson & Jubas (1995), while for the neutrinos the RHS is zero. In the second line we have expanded the $\eta$ derivative, which is formally the change along the photon path, $n^i F_{|i}$, and written it schematically in terms of $\dot{x}^i$ and $\dot{n}^i$, hence the approximately equal sign. In a flat universe the $\dot{n}^i$ term vanishes: the geodesic is a straight line. It is this term which gives rise the the $\kappa_\ell$ factors in the recurrence relation for the $G_\ell$.

We can evaluate the $\dot{q}$ term from the geodesic equation:

$$\begin{aligned} \dot{q} &= -\frac{q}{2} \dot{h}_{ij} n^i n^j \\ &= \frac{q}{3} \left( -\dot{h} Q + 2\dot{H}(1 - 3K/k^2)^{-1} G_2 \right). \end{aligned} \tag{69}$$

Putting these terms together, and using the recurrence relation (57) to rewrite the derivative along the path, we find the following (recall $\theta = i\vec{k} \cdot \vec{v}$):

$$\dot{\delta} = -\frac{4}{3}\theta - \frac{2}{3}\dot{h}$$

$$\begin{aligned}
\dot{\theta} &= k^2(\frac{1}{4}\delta - \frac{1}{10}\kappa_2 F_2) + an_e\sigma_T(\theta_B - \theta) \\
\dot{F}_2 &= \frac{8}{3}\kappa_2\theta - \frac{3}{7}k\kappa_3 F_3 + \frac{4}{3}\kappa_2^{-1}\dot{H} - \frac{9}{10}an_e\sigma_T F_2 \\
\dot{F}_\ell &= k\left[\frac{\ell}{2\ell-1}\kappa_\ell F_{\ell-1} - \frac{\ell+1}{2\ell+3}\kappa_{\ell+1}F_{\ell+1}\right] - an_e\sigma_T F_\ell,
\end{aligned} \quad (70)$$

where for the neutrinos the terms proportional to $\sigma_T$, coming from the collision term, are absent. The $\ell = 2$ mode is related to the anisotropic stress. As can be seen, for $\ell \leq 1$ the equations are formally identical to those for $K = 0$.

The modification of the equations for photon polarization (Bond & Efstathiou 1984, Kosowsky 1995, Zaldarriaga & Harari 1995) simply requires inserting factors of $\kappa_\ell$ in the mode coupling terms just as above. Since for $\Omega_{\rm tot} > 1$ the sound horizon can subtend a larger angular scale than in a flat universe, the polarization-to-temperature ratio can be larger in a closed universe, as one would expect by analogy with the open case (Zaldarriaga & Harari 1995).

The remaining (perturbed Einstein and fluid) equations are gathered here for completeness:

$$\begin{aligned}
\frac{d}{d\eta}\left(a\dot{h}\right) &= (8\pi G)a^3 \sum_i \bar{\rho}_i(1+3w_i)\delta_i \\
\frac{k^2}{3}\left(\dot{h} + \dot{H}\right) &= -(8\pi G)a^2 \sum_i \bar{\rho}_i(1+w_i)\theta_i \\
\dot{\delta}_{CDM} &= -\frac{1}{2}\dot{h} \\
\dot{\delta}_B &= -\theta_B - \frac{1}{2}\dot{h} \\
\theta_{CDM} &= 0 \\
\dot{\theta}_B &= -\frac{\dot{a}}{a}\theta_B + \frac{4\bar{\rho}_\gamma}{3\bar{\rho}_B}an_e\sigma_T(\theta_\gamma - \theta_B),
\end{aligned} \quad (71)$$

where the sum $\sum_i$ is over the components: CDM, baryons, photons, neutrinos.

The initial conditions for the metric, density and velocity perturbations are unchanged by curvature, as one would expect physically, and can be found in Ma & Bertschinger (1995). The initial value of the neutrino anisotropic stress acquires a factor of $\kappa_2$ in a closed universe. Due to our choice of normalization the expression for $C_\ell$ is the same as for the flat space case:

$$(2\ell+1)^2 C_\ell = \frac{\sqrt{K}}{8\pi}\sum_{\beta=3}^{\infty}\frac{1}{\beta}P_{\rm prim}(\beta)\,\beta^3\,|F_\ell(\beta,\eta_0)|^2, \quad (72)$$

where $\eta_0$ is the conformal time today. The factor of $\sqrt{K}$ multiplying the sum over $\beta$ makes the limit $\sqrt{K} \to 0$ smooth, and should also appear in the definition of the moments of the matter power spectrum.

The primordial power spectrum, $P_{\rm prim}(\beta)$, comes from a theory of fluctuation generation. Rather than perform calculations of the primordial spectrum for specific inflationary potentials, we have assumed that the spectrum is either a power-law in $\beta$ (with spectral index unity), or that it corresponds to constant fluctuations in the gravitational potential $\Phi$ per logarithmic interval in wavenumber (which in the flat case is the same as the previous assumption). For this latter case, generalizing (Lyth & Stewart 1990, or see White & Bunn 1995) gives

$$P_{\rm prim}(\beta) \propto \frac{(\beta^2-4)^2}{\beta(\beta^2-1)} \quad (73)$$

which we shall refer to as "scale invariant".